\documentclass[12pt,preprint]{aastex}

\newcommand{\tur}{4C~41.17}
\newcommand{\squ}{4C~60.07}
\newcommand{\hor}{B2~0902+34}
\newcommand{\boa}{MRC~1243+036}

\newcommand{\etal}{{et al.}}
\newcommand{\eg}{{e.g.,}}
\newcommand{\lya}{\ifmmode {\rm Ly\alpha}\else{\rm Ly$\alpha$}\fi}
\newcommand{\hzrgs}{HzRGs}
\newcommand{\hzrg}{HzRG}

\newcommand{\LFIR}{\ifmmode {\rm \,L_{FIR}}\else ${\rm \,L_{FIR}}$\fi}
\newcommand{\LUV}{\ifmmode {\rm \,L_{UV}}\else ${\rm \,L_{UV}}$\fi}
\newcommand{\Llya}{\ifmmode {\rm \,L_{Ly\alpha}}\else ${\rm \,L_{Ly\alpha}}$\fi}
\newcommand{\OmM}{\ifmmode {\Omega_{\rm M}}\else $\Omega_{\rm M}$\fi}
\newcommand{\OmL}{\ifmmode {\Omega_{\Lambda}}\else $\Omega_{\Lambda}$\fi}
\newcommand{\ph}{\ifmmode {h_{65}^{-1}}\else $h_{65}^{-1}$\fi}
\newcommand{\psqh}{\ifmmode {h_{65}^{-2}}\else $h_{65}^{-2}$\fi}
\newcommand{\degree}{\ifmmode {^{\,\circ}} \else {$^{\,\circ}$}\fi}
\newcommand{\Lsun}{\ifmmode {\rm\,L_\odot}\else ${\rm\,L_\odot}$\fi}
\newcommand{\Msun}{\ifmmode {\rm\,M_\odot} \else ${\rm\,M_\odot}$\fi}
\newcommand{\Zsun}{\ifmmode {\rm\,Z_\odot} \else ${\rm\,Z_\odot}$\fi}
\newcommand{\kmps}{\ifmmode {\rm\,km~s^{-1}} \else ${\rm\,km\,s^{-1}}$\fi}
\newcommand{\kpc}{{\rm\,kpc}} 
\newcommand{\ergps}{\ifmmode {\rm\,erg\,s^{-1}} \else {${\rm\,erg\,s^{-1}}$}\fi}
\newcommand{\ergpspcm}{\ifmmode {\rm\,erg\,s^{-1}\,cm^{-2}} \else {${\rm\,erg\,s^{-1}\,cm^{-2}}$}\fi}
\newcommand{\surfbr}{\ifmmode {\rm\,erg\,s^{-1}\,cm^{-2}\,arcsec^{-2}} \else {${\rm\,erg\
,s^{-1}\,cm^{-2}\,arcsec^{-2}}$}\fi}
\newcommand{\Msunpyr}{\ifmmode {\rm\,M_\odot\,yr^{-1}} \else {${\rm\,M_\odot\,yr^{-1}}$}\fi}
\newcommand{\pyr}{\ifmmode {\rm\,yr^{-1}} \else {${\rm\,yr^{-1}}$}\fi}
\newcommand{\psec}{\ifmmode {\rm\,s^{-1}} \else {${\rm\,s^{-1}}$}\fi}

\begin{document}

\title{Giant Ly$\alpha$ nebulae associated with high redshift radio galaxies\altaffilmark{1}}

\author{Michiel Reuland\altaffilmark{2,3,4}, Wil van
Breugel\altaffilmark{2}, Huub R\"ottgering\altaffilmark{3}, Wim de
Vries\altaffilmark{2}, S.A. Stanford\altaffilmark{2,4}, Arjun
Dey\altaffilmark{5}, Mark Lacy\altaffilmark{6}, Joss
Bland-Hawthorn\altaffilmark{7}, Michael Dopita\altaffilmark{8},
and George Miley\altaffilmark{3}}

\altaffiltext{1}{Based on observations made at the W.M. Keck
Observatory, which is operated as a scientific partnership among the
California Institute of Technology, the University of California and
the National Aeronautics and Space Administration. The Observatory was
made possible by the generous financial support of the W.M. Keck
Foundation.}

\altaffiltext{2}{Institute of Geophysics and Planetary Physics,
Lawrence Livermore National Laboratory, L$-$413, P.O. Box 808,
Livermore, CA 94551 USA, email: mreuland, wil@igpp.ucllnl.org}

\altaffiltext{3}{Leiden Observatory, P.O. Box 9513, 2300 RA Leiden The
Netherlands, email: rottgeri, miley@strw.leidenuniv.nl}

\altaffiltext{4}{Department of Physics, UC Davis, One Shields Avenue,
Davis, CA 95616 USA}

\altaffiltext{5}{KPNO/NOAO, 850 N. Cherry Ave., P.O. Box 26732,
Tucson, AZ 85726 USA}

\altaffiltext{6}{SIRTF Science Center, Caltech, MS 220-6, 1200
E. California Boulevard, Pasadena, CA 91125 USA}

\altaffiltext{7}{Anglo$-$Australian Observatory, P.O. Box 296, Epping,
NSW 2121 Australia}

\altaffiltext{8}{Research School of Astronomy and Astrophysics,
The Australian National University, Cotter Road, Weston Creek, ACT2611,
Australia}

\begin{abstract}

\noindent We report deep Keck narrow-band \lya\ images of the luminous
$z > 3$ radio galaxies \tur, \squ, and \hor.  The images show giant,
100$-$200\kpc\ scale emission line nebulae, centered on these
galaxies, which exhibit a wealth of morphological structure, including
extended low surface brightness emission in the outer regions,
radially directed filaments, cone-shaped structures and (indirect)
evidence for extended \lya\ absorption.  We discuss these features
within a general scenario where the nebular gas cools gravitationally
in large Cold Dark Matter (CDM) halos, forming stars and multiple
stellar systems. Merging of these ``building'' blocks triggers large
scale starbursts, forming the stellar bulges of massive radio
galaxy hosts, and feeds super-massive black holes which produce the
powerful radio jets and lobes.  The radio sources, starburst
superwinds and AGN radiation then disrupt the accretion process
limiting galaxy and black hole growth, and imprint the observed
filamentary and cone-shaped structures of the \lya\ nebulae.\\
\end{abstract}


\section{Introduction}

High redshift radio galaxies (\hzrgs; $z > 2$) are great beacons for
pinpointing some of the most massive objects in the early universe, whether
these are galaxies, super-massive black holes or even clusters of
galaxies \citep{vanBreugel02spie}.  

Powerful, non-thermal radio sources are uniqely associated with
massive, multi $L_\star$ elliptical galaxies.  At {\it low} redshifts
this has been known since the first optical identifications of
extra-galactic radio sources were made possible using radio
interferometers \citep[Cygnus~A;][]{CarilliBarthel96aarv}.  The large,
twin-jet, double-lobe morphologies and enormous radio luminosities
($P_{178\,\rm MHz} \sim 5 \times 10^{35} \rm \ergps\,Hz^{-1}$)
suggested already early on that such galaxies must have spinning,
super-massive black holes (SMBHs) in their centers
\citep{Rees78nat,BlandfordPayne82mnras}.  We now know that the masses
of the stellar bulges of galaxies and their central black holes are
closely correlated \citep[$M_{SMBH} \sim 0.006
M_{bulge}$;][]{Magorrian98aj,Gebhardt00apj,FerrareseMerritt00apj},
indicating a causal connection and it is no longer a surprise that
their parent galaxies occupy the upper end of the galaxy mass function
($\gtrsim 2 \times 10^{11}$\,\Msun).

At {\it high} redshifts the
evidence is more recent but equally compelling.  The combined
near-infrared ``Hubble'' $K-z$ relation for radio and field galaxies
\citep{DeBreuck02aj,Jarvis01mnras} shows that radio-loud galaxies are
the most luminous at any redshift $0 < z < 5.2$. This is despite
considerable changes in their rest-frame morphologies from smooth
ellipticals at $z \sim 1$ \citep{Best98mnras} to large ($\sim 50 - 70$
kpc) multi-component, structures aligned with the radio source at $z >
2$ \citep{Pentericci99aa,vanBreugel98apj}.

Evidence that the \hzrgs\ are young forming galaxies has been provided
by the direct detection of absorption lines and P Cygni-like features
from young hot stars \citep{Dey97apj} and strong sub-mm continuum
emission
\citep{Dunlop94nat,Ivison98apj,Archibald01mnras,Reuland03prepa}.  The
sub-mm observations imply ``hyper'' luminous rest-frame far-infrared
luminosities ($L_{FIR} \gtrsim 10^{13} \Lsun$) and huge ($> 1000
\Msunpyr$) star formation rates (SFRs). Large amounts of star
formation are also indicated by the recent detections in several
\hzrgs\ of very extended ($\sim 30 - 50$ kpc) molecular gas and dust
clouds \citep{Papadopoulos00apj,DeBreuck02prep}, showing that the star
formation in these systems occurs on galaxy wide scales.

There are great practical advantages in using \hzrgs\ to study the
formation and coevolution of massive galaxies and their central black
holes. Radio source identifications are unbiased with respect to heavy
obscuration by dust, which is especially important in young galaxies
with large amounts of star formation. Furthermore, \hzrgs\ are the
most extended and luminous galaxy-sized structures at many wavelengths
and can therefore be studied in great detail over nearly the entire
electromagnetic spectrum.  Some specific questions that one might hope
to address are: Do the stellar bulges of massive elliptical galaxies
form mostly during one major, ``monolithic'' collapse or over a longer
period through the merging of smaller components? Does outflow and
radiation from starburst and AGN affect the galaxy formation process?
Are active, massive forming galaxies capable of enriching the
intra-cluster media with metals and thus affect cluster evolution?

It is well established that radio galaxies at low to moderately high
redshifts ($z < 2$) often have bright, extended ($ 100 - 200 \kpc$),
emission line nebulae which are aligned with the radio sources
\citep[see][and references therein]{McCarthy93araa}. The morphologies,
kinematics and ionization of these nebulae have been studied
extensively
\citep[\eg][]{vanBreugel85apj,Baum92apj,McCarthy93araa,vanOjik97aaa,
VillarMartin99mnras,Best00bmnras,Tadhunter00mnras,Inskip02amnras}. From
this work one can conclude that (i) the gas is probably leftover
material from earlier galaxy merging events which involve at least one
gas rich galaxy, (ii) the kinematics of the gas along the radio source
axes is disturbed due to interaction with the radio lobes and jets,
(iii) the merging events may have triggered star formation and radio
AGN activity at the galaxy centers, and (iv) the gas is ionized in
part by shocks induced by the radio sources and in part by
photoionization from the AGN
\citep{VillarMartin99mnras,Tadhunter00mnras,
Best00amnras,Best00bmnras,Inskip02amnras,Inskip02bmnras}.

Few images exist of extended emission-line (\lya) nebulae at much
higher redshifts ($z > 3$). Most known nebulae are radio-{\it loud},
being associated with radio galaxies, but some are radio-{\it quiet}
\citep{Steidel00apj,Francis01apj}. Nearly all these nebulae were
imaged with 4\,m class telescopes and little morphological detail can
be discerned because of limited signal-to-noise and often moderately
poor seeing.  Higher quality images are of great interest because of
the potential diagnostics they may provide about the very early stages
of galaxy formation, and about starburst/AGN feedback and chemical
enrichment during this process.

For example, the existence of radio-quiet \lya\ nebulae without obvious
central sources of ionization and/or outflow suggests that they may be
due to accretion of cooling gas in the gravitational fields of CDM halos
\citep{Steidel00apj,Chapman01apj,Fardal01apj,Francis01apj}. Such nebulae
might be the very first stage in the formation of large galaxy, or its
building blocks, a group of smaller galaxies.  The radio-loud \lya\
nebulae could be the next phase, during which the central galaxy developes
a large scale starburst as a result of galaxy merging and a super-massive
black hole is activated. The ensuing outflow and ionizing radiation
might then heat and expel the accreting gas, while adding enriched
material from the central starburst to the mix, effectively stopping
further galaxy growth until starburst and AGN activity subside. Such a
self regulated process of galaxy growth has been invoked as a possible
explanation for the surprisingly tight SMBH/stellar bulge correlation
\citep{SilkRees98aa,HaimanRees01apj}.  Could it be that in the case
of radio galaxies, which are the most massive systems at any redshift,
this process is further aided by radio jets and lobes?

To investigate the nature of \lya\ nebulae associated with \hzrgs\ and
exploit their diagnostic value we have obtained deep narrow-band \lya\
images of three $3< z < 4$ radio galaxies using the W.M. Keck Observatory
10\,m telescopes. Here we discuss the morphologies of these nebulae,
their relationship to other pertinent imaging data (radio, infrared,
mm), and the possible implications for understanding the formation and
coevolution of massive galaxies and super-massive black holes.

Our paper is organized as follows. 
In Section 2 we describe the target selection, observations and data
analysis. The observational results for the individual objects are
presented in Section 3 and some physical parameters are deduced.  In
Section 4, we discuss possible scenarios for the origin of the outer
nebulae and the features in the central region. We summarize our
conclusions in Section 5.
Throughout we adopt a flat universe cosmology with $\OmM = 0.3$, $\OmL =
0.7$, and $H_{0} = 65 \kmps\,\rm Mpc^{-1}$.  Note that at the high
redshifts of our targets a slightly different choice of cosmological
parameters can make a significant difference to the derived
parameters. To emphasize this we quote our numbers while retaining the
scale factor $h_{65} = 1$. At the redshifts of our galaxies ($z = 3.4
-3.8$) and using the adopted cosmology we find look-back times
$\sim$12.8\,\ph\,Gyr, galaxy ages $\lesssim$1.7\,\ph\, Gyr, and
angular scales of $\sim$7.6\,\ph\,$\rm \kpc~arcsec^{-1}$.

\section{Sample Selection and Observations}

The observations presented here include sensitive optical narrow-band
and broad-band imaging of three $z > 3$ radio galaxies using the Keck
telescopes and previously unpublished Hubble Space Telescope
(HST) imaging of \squ. Our targets were selected on the basis of their 
high redshifts and the availability of data at many other wavelengths.

\subsection{Sample Selection}

\centerline{\it \tur} \tur\ at $z = 3.798$ was chosen because its
\lya\ halo is one of the most luminous and extended known. Previous
imaging observations showed a bright elliptical-shaped \lya\ halo with
a major axis of $\sim 15$\arcsec\ directed along the radio axis
\citep{Chambers90apj,Hippelein93nat,Adam97aa,RoccaVolmerange99amsproc},
and spectroscopic evidence indicated that the \lya\ might extend to
$\gtrsim 18\arcsec$ \citep{Dey99amsproc}.  The radio source has a
multiple component, asymmetric, ``double-double'' FR II morphology,
with the radio axis of the inner source having a position angle
different from the outer source \citep{Chambers90apj,Carilli94aj}.  It
was the first \hzrg\ in which a large dust mass has been detected
through sub-mm observations \citep{Dunlop94nat}, implying a large SFR
in the range of $2000 - 3000 \Msunpyr$.

The galaxy has a clumpy rest-frame UV morphology, with
the brightest components aligned with the inner radio source
\citep{vanBreugel99amsproc}.  This light is unpolarized between
$\lambda_{\rm rest} \sim 1400 - 2000$\,\AA, and thus not due to
scattering from a hidden quasar-like AGN \citep{Dey97apj}.  Instead,
the observations show absorption line features from young hot stars which
resemble those seen in $z \approx 2-3$ star-forming galaxies and nearby
starburst systems.  Dey \etal\ derive a SFR of $140 - 1100 \Msunpyr$ for
the central $10\kpc \times 20\kpc$ of \tur\, or $400 - 3200 \Msunpyr$
after correction for local extinction. These values are consistent
with those derived from the sub-mm observations. In this radio source
aligned star formation is thought to have been triggered by radio jets
colliding with a large, dense cloud in the forming galaxy \citep{Dey97apj,
vanBreugel99amsproc,Bicknell00apj}.

\smallskip
\centerline{\it \squ} \squ\ \citep[$z = 3.791$;][]{Chambers96apj} is
close to the Galactic plane (b$_{\rm II}$ = 12\degree) and suffers
significant Galactic extinction (see, Table \ref{obstable}).  Despite
this unfortunate circumstance, we selected \squ\ as one of our targets
because its redshift is close to that of \tur\ and it is a strong
sub-mm emitter and one of only three \hzrgs\ in which CO has been
detected \citep{Papadopoulos00apj,DeBreuck02prep}.  The large extent
(7\arcsec\ $\sim 50\ph \kpc$) and dynamical mass ($\sim 1-7 \times
10^{11} \Msun$) of the cold gas are some of the best evidence that
\squ\ is indeed a massive forming galaxy.  A low signal-to-noise image
of \squ\ shows a 2\arcsec\ \lya\ feature aligned with the radio
source.  The source displays a simple edge brightened FR II
morphology \citep{FanaroffRiley74mnras}, unlike the other two objects
in our sample \citep{Carilli94aj,Chambers96apj}.

\smallskip
\centerline{\it \hor} Our third target, \hor\ \citep[$z =
3.395$;][]{Lilly88apj}, was selected because of its extended, very
diffuse optical morphology \citep{EisenhardtDickinson92apj}, which is
very unlike that of the elongated, radio source aligned structures
seen in \tur\ and \squ. It is also the only \hzrg\ for which 21\,cm
neutral hydrogen has been detected in absorption against the radio
continuum \citep{Uson91,Briggs93apj}. No other evidence for cold gas
or dust has been found in \hor. No strong absorption in the \lya\
emission has been detected through spectroscopy
\citep{MartinMirones95apj} and no significant sub-mm signatures of
thermal dust emission have been found
\citep{Archibald01mnras}. \citet{EisenhardtDickinson92apj} published a
shallow \lya\ image for \hor\, that showed its halo to be relatively
bright but did not reveal much detail. They argued that it is a true
protogalaxy, based on its relatively flat UV-optical spectral energy
distribution, indicative of a dominant population of young stars.  The
radio structure resembles that of quasars with a flat spectrum
nucleus, a bright knotty jet with a sharp 90\degree\ bend in the
north, and a disembodied double hotspot to the
south. \citet{Carilli95aa} explained its unusual properties, inferring
that the radio axis is close to the line of sight.

\subsection{Keck Imaging}

\subsubsection{Ly$\alpha$ imaging}

We observed \tur, \squ\ and \hor\ using two custom-made, high
throughput narrow-band filters. The central wavelengths and bandpasses
were chosen to cover redshifted \lya\ within a velocity range of
$\pm$1500\kmps\ for each object, which is the typical maximum width of
the emission lines of \hzrgs\ \citep[\eg][]{Dey97apj,vanOjik97aaa}.
\tur\ and \squ\ are close in redshift and fit in the same filter,
although for \squ\ emission at velocities larger than 1000\kmps\
blueward from the systemic velocity falls outside the bandpass. The
transmission curve of this filter ($\lambda_{\rm c} = 5839.0$\,\AA\
and $\Delta \lambda = 65.0$\,\AA) averages $\sim 84$\% over the
bandpass. The narrow-band filter for \hor\ has $\lambda_{\rm c} =
5342.4$\,\AA\ and a FWHM of $\Delta \lambda = 56.8$\,\AA, with an
average transmission of 73\%.  The fields of view through the 10\,cm
$\times$ 10\,cm filters were vignetted slightly, resulting in an
useful sky coverage of $\sim 2\farcm0 \times 1\farcm7$.

The observations were made during the nights of UT 2000 December 27,
28 and UT 2001 February 25 using the Echelle Spectrograph and Imager
\citep[ESI;][]{Sheinis00spieproc} at the Cassegrain focus of the Keck II
10\,m telescope in imaging mode.  The detector used is a high-resistivity
MIT-Lincoln Labs 2048 $\times$ 4096 CCD, has a plate scale of 0\farcs154
pixel$^{-1}$, and is only partly illuminated in imaging mode.  Exposures
were broken into integrations of 1200 seconds and we performed 15\arcsec\
offsets between each integration. This facilitated removal of cosmic
rays and bad pixels on the CCD. The data were taken during photometric
conditions and good seeing (FWHM $0\farcs6 - 0\farcs8$ in the co-added
images) and were reduced using standard methods in IRAF (including
registering and stacking the individual integrations). Flatfielding was
done with twilight sky flats only (as opposed to creating field flats from
unregistered images) to prevent suppression of faint diffuse emission.

The data were flux calibrated using the spectrophotometric standard stars
Feige~34, PG~1121+145, and PG~1545+035 \citep{Massey88apj}. All magnitudes
and colors quoted in this paper are on the AB system and the individual
calibrations agree to within approximately 10\%, which we take to be the
systematic photometric uncertainty.  We corrected for Galactic extinction
using the E(B$-$V) values based upon IRAS 100$\mu$m cirrus emission maps
\citep{Schlegel98apj} and extrapolating following \citet{Cardelli89apj}.
A summary of the total integration times, surface brightness limits,
and Galactic extinction corrections is given in Table \ref{obstable}.
With 1$\sigma$ surface brightness limits of NB $\sim 29  - 30$ mag within
an aperture of the seeing disk ($\sim 0.7\arcsec$) these are the deepest
narrow-band exposures of \lya\ nebulae ever obtained.

\subsubsection{Broad-band imaging}

We also obtained emission line free $R$-band images during the same
nights to allow continuum subtraction from the narrow-band exposures
and create pure \lya\ images.  For \tur\ and \squ\ the $R$-band
($\lambda_{c} = 6700$\,\AA; FWHM = 1400\,\AA) is just redward of \lya\
and free of strong emission lines \citep[all emission lines that fall
within the filter bandpass have observed equivalent widths $\lesssim$
10\,\AA;][Reuland \etal\ in preparation]{Dey97apj}. In the case of \hor\ the
$R$-band contains the \ion{C}{4} $\lambda\lambda 1548,1551$ doublet
and \ion{He}{2} $\lambda 1640$.  The spectrum by
\citet{MartinMirones95apj} shows that \ion{C}{4} and \ion{He}{2} have
observed equivalent widths of $\sim$ 55\,\AA, and $\sim$ 40\,\AA,
respectively. The average total contribution of these emission lines
to the total flux received in the broad-band filter is $\lesssim 7$\%
indicating that they are only minor contaminants. However, the
relative contribution could depend on the region of interest. There is
no easy way to correct for this, but compared to the large equivalent
width \lya\ line continuum subtraction is only a small correction and
does not change any morphological features of interest for this paper.
These $R$-band images have fields of view of $3\farcm3 \times
1\farcm7$, were broken into 400\,s exposures and were reduced
following standard methods.

\subsection{HST Imaging}

\squ\ was observed during Cycle 6 with the Wide Field Planetary Camera
2 \citep[WFPC2;][]{Trauger94apj} on the HST. The object was placed on
the PC chip, which utilizes an 800 $\times$ 800 pixel Loral CCD as
detector with a plate scale of 0\farcs0455 pixel$^{-1}$. In two
pointings 9 broad-band exposures of 2900\,s each were obtained for a
total exposure time of 26.1\,ks through the F702W filter, which has a
central wavelength $\lambda_{\rm c} = 6944.3$\,\AA\ and a FWHM of
$\Delta \lambda = 1384.7$\,\AA.  For \squ\ this filter includes the
\ion{C}{4} doublet with a combined observed equivalent width of $\sim
300$\,\AA\ (Reuland \etal\ in preparation), but is free from other strong
emission lines.  After cosmic ray removal these images were
co-added. The data were severely affected by light scattering off of
the Earth atmosphere. However, the scattered light pattern was fixed
with respect to the CCD. This allowed for using the small offsets
between the unregistered images of the two pointings to construct a
scattering model. After subtraction of this model, the resulting image
was virtually fully corrected.

Deep (21.6\,ks) F702W observations of \tur\ were obtained using the PC
section of WFPC2 \citep[for details
see][]{vanBreugel99amsproc,Bicknell00apj}.  Observations with the
F702W filter are slightly contaminated by the \ion{C}{4} doublet,
which has a combined observed equivalent width of $\sim 104$\,\AA\
\citep{Dey97apj}.

Eisenhardt and Dickinson observed \hor\ during Cycle 4 for 21.6\,ks
using the PC of WFPC2 with the filter F622W, which is centered at
$\lambda_{\rm c} = 6189.6$\,\AA, has a FWHM of $\Delta \lambda =
917.1$\,\AA, and is free of emission from the \lya\ and \ion{C}{4}
lines. These data were obtained from the archives.

\subsection{Relative Astrometry}

Astrometric calibration of the Keck images was performed using the
USNO-A2.0 catalog \citep{Monet98aas}. For the broad-band images 15
catalog stars were typically available in each field, of which about
10 were unsaturated and suitable for astrometry. In all cases this
resulted in astrometry accurate to approximately 0\farcs15 rms with
respect to the USNO catalog. The catalog itself has a rms uncertainty of
$\sim 0\farcs25$ with respect to the International Celestial Reference
Frame (ICRF), while the astrometric uncertainties of the radio images
are $\lesssim 0\farcs2$ with respect to the ICRF \citep[for a concise
discussion of relative astrometric uncertainties see][and references
therein]{DeBreuck02aj}. The narrow-band image of \hor\ contains only
few stars and the astrometry had to be bootstrapped from the broad-band
image which has more stars in the field. To do this, the narrow-band
image was registered relative to the $R$-band image to better than a
fraction of a pixel using a second order polynomial function. Similarly,
the HST images were matched to the respective Keck images resulting in
relative optical astrometry better than 0\farcs1 rms, and astrometry
relative to the radio maps of comparable accuracy as for the Keck imaging.

Generally, for the 5\,GHz and 8\,GHz VLA radio images
\citep{Carilli94aj, Carilli95aa, Carilli97apj}, we used the astrometry
as given by the observations of the phase calibrators. However, in the
case of \tur\ there was a faint optical counterpart with a well
defined centroid to the serendipitous source to the SE in the 5\,GHz
map \citep[Source E in][]{Chambers96apj}. We used this source to
register the radio and optical images to better than 0\farcs1. A
change of only $+$0\farcs15 in right ascension and $+$0\farcs2 in
declination was required to align the images. These shifts agree with
typical uncertainties in the match between the optical and radio
reference frames as cited above.  Therefore, we believe that,
generally, relative astrometry between the radio and optical is better
than $\lesssim 0\farcs35$ rms, while it is better than 0\farcs1 rms in
the case of \tur.

\subsection{Continuum subtraction}

Pure emission line images were constructed as follows: First, the
point spread functions (PSFs) of the narrow-band images were matched
to the broader PSFs of the $R$-band images. For this we selected a
suitable star close to the galaxy in both images and calculated a
convolution kernel using the IRAF task {\tt psfmatch}. Subsequently,
the \lya\ image was convolved with this kernel to obtain a new image
with a PSF similar to the $R$-band image.
 
The registered $R$-band images were then divided by a scaling factor,
and subtracted from the narrow-band images in order to remove
continuum emission. This scaling factor was determined for each radio
galaxy by convolving high quality spectra \citep[][Reuland \etal\ in
preparation]{Dey97apj}, after removal of the \lya\ emission line, with
the $R$-band and narrow-band filter curves. For each galaxy this
resulted in an expected ratio of continuum flux densities within both
filters. These expected ratios are based only on the filter curves,
while telescope and CCD efficiencies should be taken into account as
well. We corrected for this by adjusting the scaling factor by $\sim
10\%$, which was the difference between the predicted and observed
ratios for the spectrophotometric standards.

\section{Results}

Our narrow-band images show unprecedented detail in the spectacular
nebulae that surround the radio galaxies.  The nebulae extend over
more than 100\kpc\ and display complex morphologies such as
filamentary structures, conical shapes, and distinct regions of high
and low intensity. Grayscale representations (colorcycled to show both
low and high surface-brightness regions) of the pure emission line
images and contour plots of the narrow-band images (including
continuum) are shown in Figures 1$-$8. In the following, we discuss
the morphological features of the nebulae in detail. A summary of
their global properties is given in Table \ref{proptable}. This table
also gives estimates for the mass of the ionized gas and star
formation rates.  Despite being subject to considerable uncertainties
these might help provide some insight in the nature of the nebulae.

\subsection{\tur} 

\begin{figure}[t]
\plotone{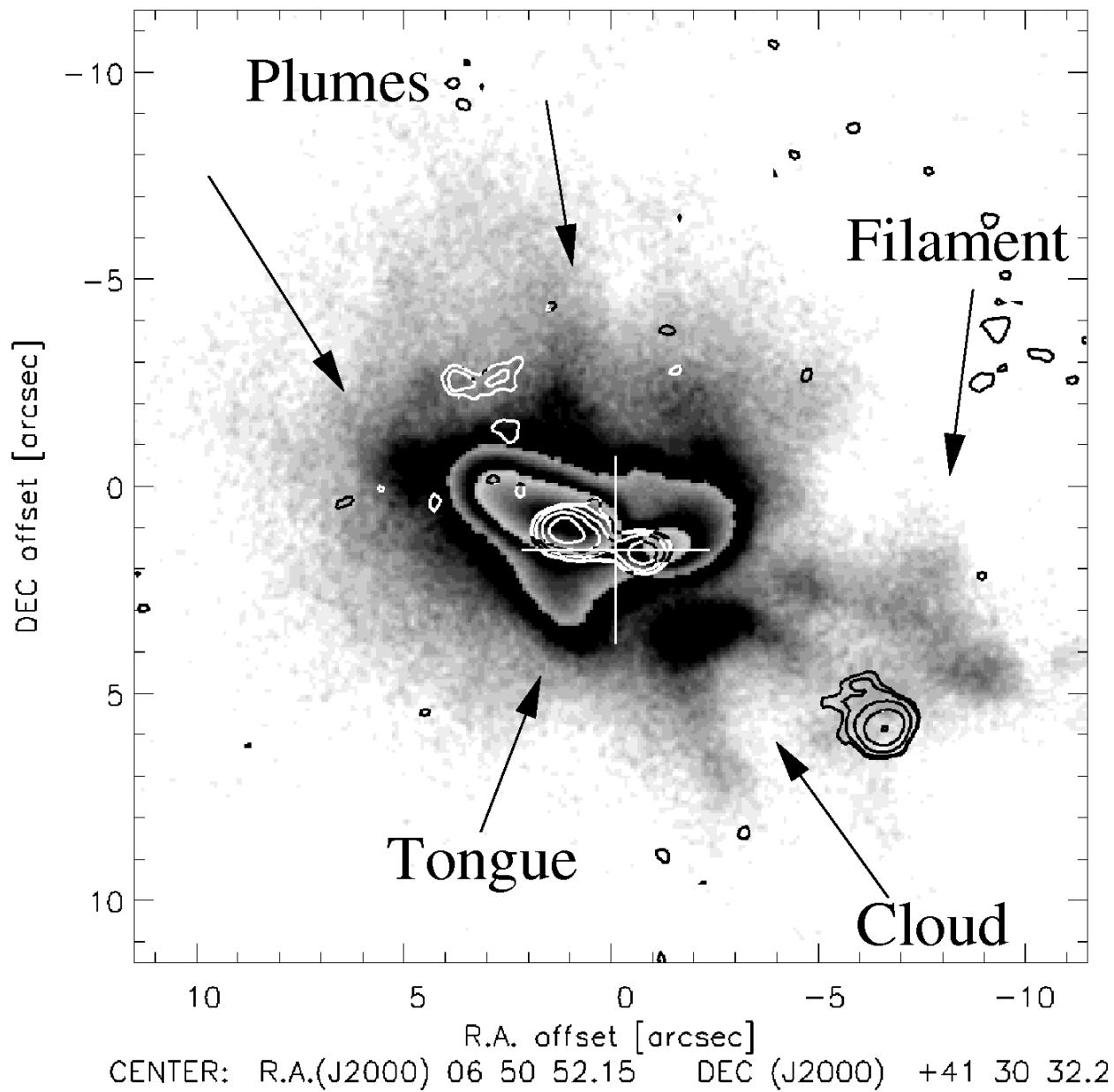}
\caption{Grayscale \lya\ image of \tur\ with a contour representation
of the 4.9\,GHz VLA map \citep{Carilli94aj} overlaid. The grayscale
has been colorcycled to show the details of the high and low surface
brightness simultaneously. The radio core is identified with a cross,
and the contour levels are 0.07, 0.11, 0.4, 1.6 and 6.4\,mJy
beam$^{-1}$. The arrows indicate ``plumes'' of enhanced emission on
both sides of the Northern Lobe, a separate emission line cloud with
filaments extending to the SSW and SW, a large filament, and a Lya
tongue corresponding to a region which appears unaffected by the radio
source.}\label{4c41Lya}
\end{figure}

\begin{figure}[t]
\plotone{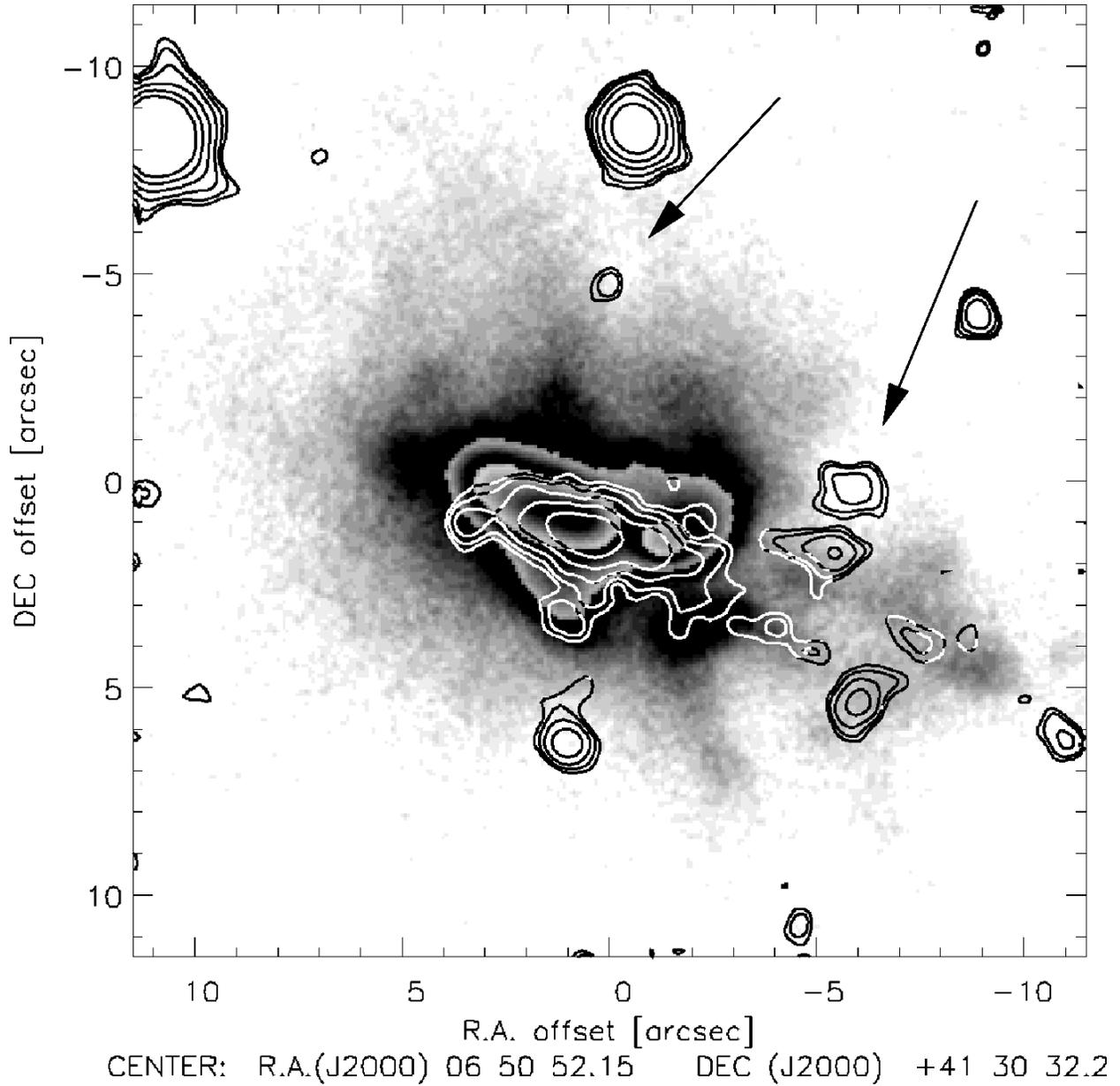}
\caption{Similar to Figure \ref{4c41Lya} but with $K$-band image
\citep[contours;][]{Graham94apj} overlaid. The contours are at
arbitrary levels and have been chosen to show the desired components.
Some dips in the outer halo are co-spatial with $K$-band galaxies
(indicated with arrows), suggesting that these galaxies are absorbing
systems close in redshift to \tur.}\label{4c41K}
\end{figure}

\begin{figure}[t]
\plotone{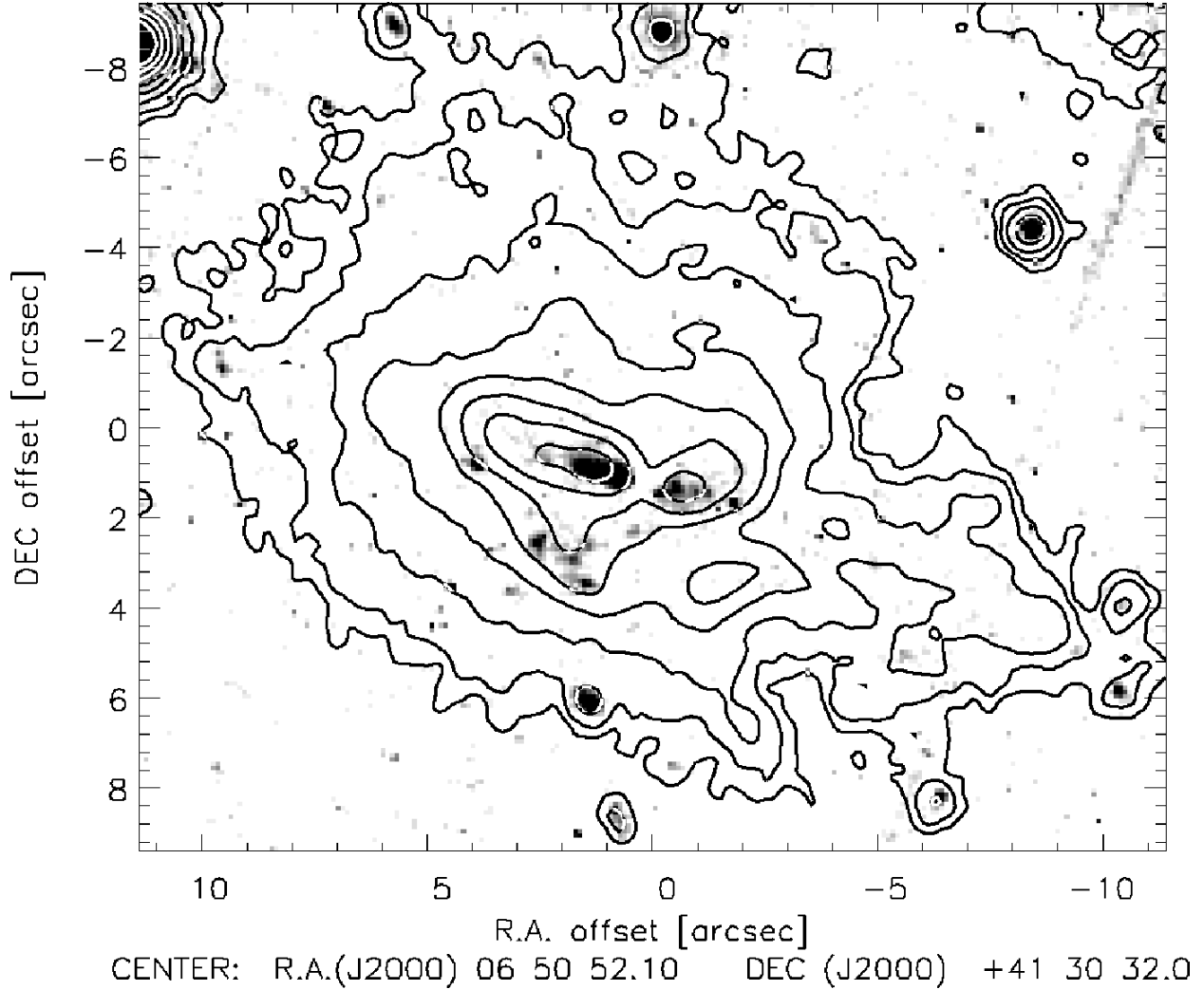}
\caption{HST F702W image \citep{vanBreugel99amsproc} of \tur\ with
narrow-band image (contours; including continuum) overlaid. The
contours indicate observed surface brightness levels at $6.7 \times
10^{-19} \times (6, 12, 25, 50, 100, 200, 400, 800) \ergpspcm\,\rm
arcsec^{-2}$. Note the group of kpc-sized clumps in the extended
diffuse emission tongue 2\arcsec\ south of the central peak, which is
not aligned with the radio axis.}\label{4c41HST}
\end{figure}

The co-added continuum-subtracted narrow-band image for \tur\ overlaid
with a 5\,GHz VLA radio map \citep{Carilli94aj} is shown in Figure
\ref{4c41Lya}, while Figure \ref{4c41K} shows the same image overlaid
with a Keck $K$-band image \citep{Graham94apj}. Figure \ref{4c41HST}
shows an overlay of the HST image with the narrow-band image (including
continuum). The \lya\ nebula is seen to extend over 25\arcsec $\times$
17\arcsec\ ($\sim 190\ph\kpc \times 130\ph\kpc$), nearly twice the
size previously seen in the images obtained with 4m class telescopes,
and shows a wealth of structure.

\noindent {\it Central region}\\ The bright inner region was observed
and discussed in detail by \citet{Chambers90apj}. It is aligned with
the inner radio source and of approximately the same size, while the
radio core is located in a central dip in the \lya\ emission.  This
has been noticed before \citep{Hippelein93nat,Adam97aa}, and
\citet{Hippelein93nat} interpret the dip as absorption by a foreground
Lyman-forest cloud unrelated to the radio galaxy. However, as proposed
by \citet{Dunlop94nat}, it seems more likely that the absorption is
due to a dust lane orthogonal to the major axis of the galaxy and its
radio source, obscuring the center of the galaxy near the position of
the radio core. This would provide a natural explanation for both the
observed sub-mm emission and the gap in the UV continuum seen in the
HST image (Fig.  \ref{4c41HST}). \citet{Graham94apj} also mention that
their $K$-band image (Fig. \ref{4c41K}; strongly contaminated with
[\ion{O}{3}] and H$\beta$) shows evidence for a double peaked
structure. They could not ascertain whether this was also true for a
line-free $Ks$ image.

\noindent {\it Galaxy scale}\\ The Keck $K$-band and HST F702W
``R''-band images (rest-frame B-band and 1500\AA\ respectively) show a
multi-component galaxy spread out over a region of 9\arcsec\ $\times$
5\arcsec\ ($\sim 68\ph\kpc \times 38\ph\kpc$) near the center of the
\lya\ nebula. A \lya\ tongue sticks out just SE of the nucleus and
overlaps with continuum emission seen in the Keck and HST images.  The
HST image shows that this is a small group of kpc-sized clumps,
embedded in low surface brightness emission of $\sim 12$\ph\kpc\
diameter, which appears to be unaffected by the radio source. From the
total UV$_{rest}$ continuum for this region
\citet{vanBreugel99amsproc} derive a star formation rate of $33
\Msunpyr$ (for the cosmological parameters used in the present
paper). For this same region we derive from our Keck observations an
integrated \lya\ flux of $3 \times 10^{-16}$ \ergpspcm, or $L_{\lya} =
5 \times 10^{43}$ \ergps.

We can convert the \lya\ luminosity into an estimate of the star
formation rate if we make a few assumptions.  With the
assumption that all the \lya\ is due to Case B recombination of
photoionization by stars we obtain $SFR \geq 8.12 \times 10^{-43}
f_{\rm esc}^{-1} L_{\lya}$.  This assumption, while unlikely to be
correct for most of the halo, may be reasonable for the undisturbed
region.  The estimated $SFR$ constitutes a lower limit as it assumes
that a fraction $f_{a} = 1$ of the ionizing photons will be
absorbed. The typical escape fraction found locally is $f_{\rm esc}
\leq 0.1$ \citep{Leitherer95apj}. However for the $z \sim 3$ Lyman
break galaxy (LBG) population \citep{Steidel96apj} $f_{\rm esc} \sim
0.5 - 1.0$ is typically found \citep{Steidel01apj} and these LBG
values may be more appropriate for \hzrgs.

From the inferred $L_{\lya}$ we obtain a lower limit to the star
formation rate of $40 f_{esc}^{-1} \Msunpyr$, rather similar to the
estimate based on the UV continuum.  However, both the UV-continuum
and \lya\ based SFR estimates are probably lower limits since they
assume no extinction, which may very well be important given the
detection of dust in \tur.

\noindent {\it Outer region and filamentary structure}\\ Perhaps the
most striking morphological feature of the \tur\ \lya\ nebula is the
cone-shaped structure emanating from the center of the galaxy, with
long filaments and a crescent-shaped cloud with horns.  The SW outer
radio lobe is located within the cone and extends further from the
core than its, much fainter, NE counterpart. It suggests that the SW
radio lobe may have encountered less dense material on its way
out. This also agrees with the emission-line and radio surface
brightness asymmetry near the center: the line emission and radio
emission of the inner radio source are brighter on the NE side. Such
optical/radio asymmetry correlations have also been seen in nearby
radio galaxies \citep{McCarthy95apj}. They are thought to be due to
local gas density asymmetries, with denser gas blocking radio jets,
resulting in brighter line and radio emission.

The filamentary/plume features of the nebula are probably caused by
AGN acivity and/or the radio source: the long SW filament ($\sim
8\arcsec$, 60 \ph \kpc) is about the same size as the distance between
the nucleus and the SW hotspot, the NE lobe appears to be embraced by two
short ``plumes'' of enhanced \lya\ emission, and the apex of the two
$\sim 4\arcsec$ (30 \ph \kpc) SW cloud horns projects close to the
nucleus.

Another point of interest is the {\it absence} of \lya\ emission near
several faint $K$-band companions around \tur\
\citep[Fig. \ref{4c41K};][]{Graham94apj}. Based on their red colors
($R - K_s \sim 4 - 5$) Graham \etal\ concluded that these are probably
multiple $L_\star$ galaxies at the same redshift as \tur. Graham's
objects 4, 9 and 12 are all near \lya\ ``gaps''.  It suggests that
they are dusty galaxies in the local foreground to the \lya\ nebula,
absorbing the \lya\ photons along the line of sight. If the sizes of
the gaps are larger than the $K$-band objects this could be evidence
for galactic envelopes \citep[\eg][]{Chen01apj}.  This is hard to
test, given the limited signal-to-noise of the $K$-band objects, but
visual inspection of Figure \ref{4c41K} shows that they are consistent
with having similar sizes. The scale of the \lya\ absorption halos
($15 \ph \kpc < D < 25 \ph \kpc$) suggests that these represent
objects which have already collapsed to ``normal'' galaxian dimensions
by this redshift.

The \tur\ field also shows an over-density of dusty galaxies on a much
larger scale \citep[$\sim 2.5\arcmin$ diameter, $\sim 1$
\ph\,Mpc;][]{Ivison00apj}, suggesting that \tur\ may be at the center
of a ``proto-cluster''.

\subsection{\squ} 

In Figure \ref{4c60Lya} we show the continuum-subtracted image of
\squ\ overlaid with the 5 GHz VLA radio map of \citet{Carilli97apj}.
Contour overlays of the narrow-band image and 1.25\,mm Plateau de
Bure map \citep{Papadopoulos00apj} on top of the newly obtained HST
image are shown in Figure \ref{4c60HST}. A zoomed-in version of this
HST image with a high resolution 8\,GHz VLA map is shown separately in
Figure \ref{4c60HSTzoom} to better bring out the morphological
details.

\noindent {\it Central region}\\ The radio core appears to fall into a
gap in the bright central \lya\ emission, and is possibly coincident
with a lack of UV continuum emission in the HST images
(Figs. \ref{4c60HST},\ref{4c60HSTzoom}) similar to \tur.  There is
enhanced emission on either side of the nucleus, and the \lya\
emission is brightest between the nucleus and the eastern lobe which
is closest to the nucleus. This emission-line/radio asymmetry
correlation is again similar to \tur\ and other radio galaxies.

\noindent {\it Galaxy scale}\\ The HST images shown in Figures
\ref{4c60HST} and \ref{4c60HSTzoom} show a string of knots subtending at
least 5\arcsec ($\sim 38\ph\kpc$). The two brightest knots appear to be on
either side of the radio core. There is a prominent Z-shaped structure,
both ends of which approximately point toward the radio lobes, making
it a typical case of the radio-optical alignment effect. By comparing
the HST morphology with the \lya\ emission, it seems that the Z-shaped
structure goes around the western peak of the halo (Fig. \ref{4c60HST}).

\noindent {\it Outer region and filamentary structure}\\ On a larger
scale the halo extends over approximately 9\arcsec\ $\times$ 7\arcsec\
($\sim 68\ph \kpc \times 53\ph \kpc$) and shows a cone shaped
structure bounded on one side by a 10\arcsec\ ($\sim 76\ph \kpc$) long
filament.  The maximum projected sizes of the \lya\ and radio
structures are fairly similar, although the filament extends beyond
the radio hotspot by 4\arcsec\ (30\,\ph\,\kpc).  The tip of the
filament is co-spatial with a galaxy (Fig. \ref{4c60HST}). Although it
is tempting to infer a connection between the two, spectroscopic
observations show that the galaxy is foreground at $z=0.891$ (Reuland
\etal\ in preparation) and cannot be related. There is also a hint of a
filament extending out to the SE.

Interestingly, the mm emission shown in Figure \ref{4c60HST} seems
almost completely anti-correlated with the \lya. The \lya\ cone and
filament appear nearly orthogonal to the cold gas and dust. This
morphology seems indicative of an outflow as it resembles luminous
starburst galaxies such as M~82.  In starburst galaxies this is
usually interpreted as being caused by a starburst driven superwind
which blows out stellar debris and interstellar gas and dust from the
central part of the galaxy \citep[\eg\
][]{StricklandStevens00mnras,Ohyama02astroph}.

\subsection{\hor} 

Figure \ref{b20902Lya} shows the \lya\ image and radio map overlay of
\hor. The HST image with narrow-band contours is shown in Figure
\ref{b20902HST}. 

\noindent {\it Central region}\\ Unlike in \tur\ and \squ\ the central
part of the \lya\ and continuum emission in \hor\ is not dominated by
a high surface brightness elongated, radio-aligned structure. Instead
the overall morphology appears more diffuse and bimodal, surrounding
the radio core and a curved jet to the north \citep[consistent
with][]{Carilli95aa,Pentericci99aa,Fabian02mnras}.  Figure
\ref{b20902HST} shows that while there is faint continuum emission to
the west co-spatial with the lower luminosity \lya\ peak, there seems
to be little continuum emission associated with the brightest peak.
Dust is not likely to be an important factor in obscuring any
continuum on the E, because the \lya\ should then be completely
extinguished in contrast to the observations and because
\citet{Archibald01mnras} did not find strong signatures of thermal
dust emission in the sub-mm. Therefore, the source of ionization in
this region remains somewhat unclear, reminiscent of the radio-quiet
\lya\ halos but with the important distinction that there is a
luminous AGN nearby. Given that the radio axis of the source is
oriented close to the line of sight \citep{Carilli95aa}, beaming
effects are expected to be important.  A possibility is that the
bright \lya\ is due to scattered light from the AGN or the result of
collisional excitation. Based on the relatively flat UV-optical
spectral energy distribution, \citet{EisenhardtDickinson92apj} argued
for the presence of a large population of young stars.  Both the inner
\lya\ morphology and the HST continuum might be better understood as
being due to a shocked cocoon of gas and possibly shock-induced star
formation associated with the approaching radio lobe.

\noindent {\it Outer region and filamentary structure}\\ The extended
emission line region subtends approximately 10\arcsec\ $\times$
8\arcsec\ on the sky ($\sim 80\ph \kpc \times 64\ph \kpc$).  Again the
radio and \lya\ structures are roughly comparable in size, and the
brightest radio and line emission are on the same side of the core.
However, the radio emission near bright \lya\ emission is strongly
polarized, unlike \tur\ and \squ.  This could be understood if the
radio jet in \hor\ is pointed towards us.  The shorter line of sight
through the gaseous medium would cause less depolarization \citep[the
Laing-Garrington effect;][]{GarringtonConway91mnras}, and at the same
time would boost the core and jet emission due to relativistic
beaming.

There are few signs of large scale filamentary emission in \hor. They
could be hidden if the filaments preferentially follow the radio axis
and are similarly elongated along our line of sight. 

\section{Discussion}

Our deep, sub-arcsecond seeing narrow-band imaging observations show
previously unknown, complex morphologies in giant \lya\ nebulae around
three \hzrgs.  Since the nebulae are centered on masssive forming
galaxies these structures may provide new insights about this galaxy
formation process and the importance of starburst/AGN feedback.  We
will discuss some of the general conclusions that can be drawn from
these data.  A more in-depth understanding of the nature of the
nebulae will require analysis of spectroscopic observations, which
will be presented in a following paper.  For now we note that these
and other spectroscopic evidence for \hzrgs\ show that the nebulae, at
least along the {\it major} axes of the radio sources, are enriched
and ionized i.e. are not composed of pure primordial (H, He) gas and
not due to \lya\ scattering off cold gas (HI). This is based on the
detection of C and O along the radio sources over many tens of kpc
\citep[e.g.,][]{Overzier01aa,Maxfield02aj,vanBreugel02spie,
Jarvis02mnras,VillarMartin02mnras}, including evidence for enriched
material \ion{C}{4} in undisturbed regions {\it beyond} radio hotspots
\citep{Maxfield02aj,VillarMartin02mnras}.

Some of the properties of radio galaxy \lya\ nebulae, to the extent
that they could be studied with mostly 4\,m class telescopes, have
been discussed by others
\citep[e.g.,][]{Chambers90apj,EisenhardtDickinson92apj,Hippelein93nat,
McCarthy93araa,vanOjik96aa,vanOjik97aaa,RoccaVolmerange99amsproc,Venemans02apj}.
Here we will focus on the newly discovered features in our images: the
very large low surface brightness outer regions, the long radial
filaments and cone-shaped structures, and the (indirect) evidence for
extended \lya\ absorption.  It is important to reiterate that
radio-{\it quiet} \lya\ nebulae also exist
\citep{Steidel00apj,Francis01apj}. It is not known whether the gas in
these nebulae is enriched or not, and the available images are of
insufficient detail to determine whether they exhibit similar
filamentary structures as the \hzrg\ nebulae.

A plausible scenario for explaining the origin of \lya\ nebulae is one
in which cold gas from the ``Dark Ages'' is accreted in large Cold
Dark Matter halos. In contrast to the classical picture of gas cooling
from the virial temperature, $T \sim 10^{6}$\,K for a typical dark
matter halo, recent theoretical models predict that most of the
infalling gas may not heat to virial temperatures but remains at $T
\sim 10^{4}$\,K because of efficient \lya\ cooling
\citep{Fardal01apj,Haiman00apj}.  There is strong evidence that
\hzrgs\ reside in (proto)-clusters
\citep[e.g.,][]{Pentericci00aa,Venemans02apj}, and a viable scenario
to explain these nebulae might then be that they are signatures of
cooling flows.

As the gas accretes it will condense into stars and galaxies, and thus
cooling flow and star formation are intimately linked.  The \lya\
emission associated with the release of gravitational potential energy
is expected to be more spatially extended than emission related to the
starburst and they are expected to be of similar magnitude in the most
luminous objects \citep{Fardal01apj}.  The fairly smooth and very
extended outer regions orthogonal to the radio sources suggest that
these could be the remnants of these initial, undisturbed accretion
flows. The surface brightness profile in the outer parts of the SE
corner of \tur, which appears to be the least disturbed part of the
nebula, drops of with radius as approximately $I(r) \propto r^{-2}$
and would be consistent with theoretical predictions for cooling halos
\citep{Haiman00apj}.

What about the radial filamentary and cone-shaped structures?  How did
these form in such a scenario?  Here some lessons might be learned
from nearby cooling flow clusters, Abell~1795 \citep{Fabian01mnras}
and NGC~1275 \citep{Conselice01aj,Fabian01mnras}, starburst superwinds
such as in M~82, Arp~220 \citep[\eg][]{Heckman90apj}, and Seyfert II's
\citep[NGC 1068;][]{Dopita02astroph}.

\subsection {Cooling flows and radio lobes}

\begin{figure}[t]
\plotone{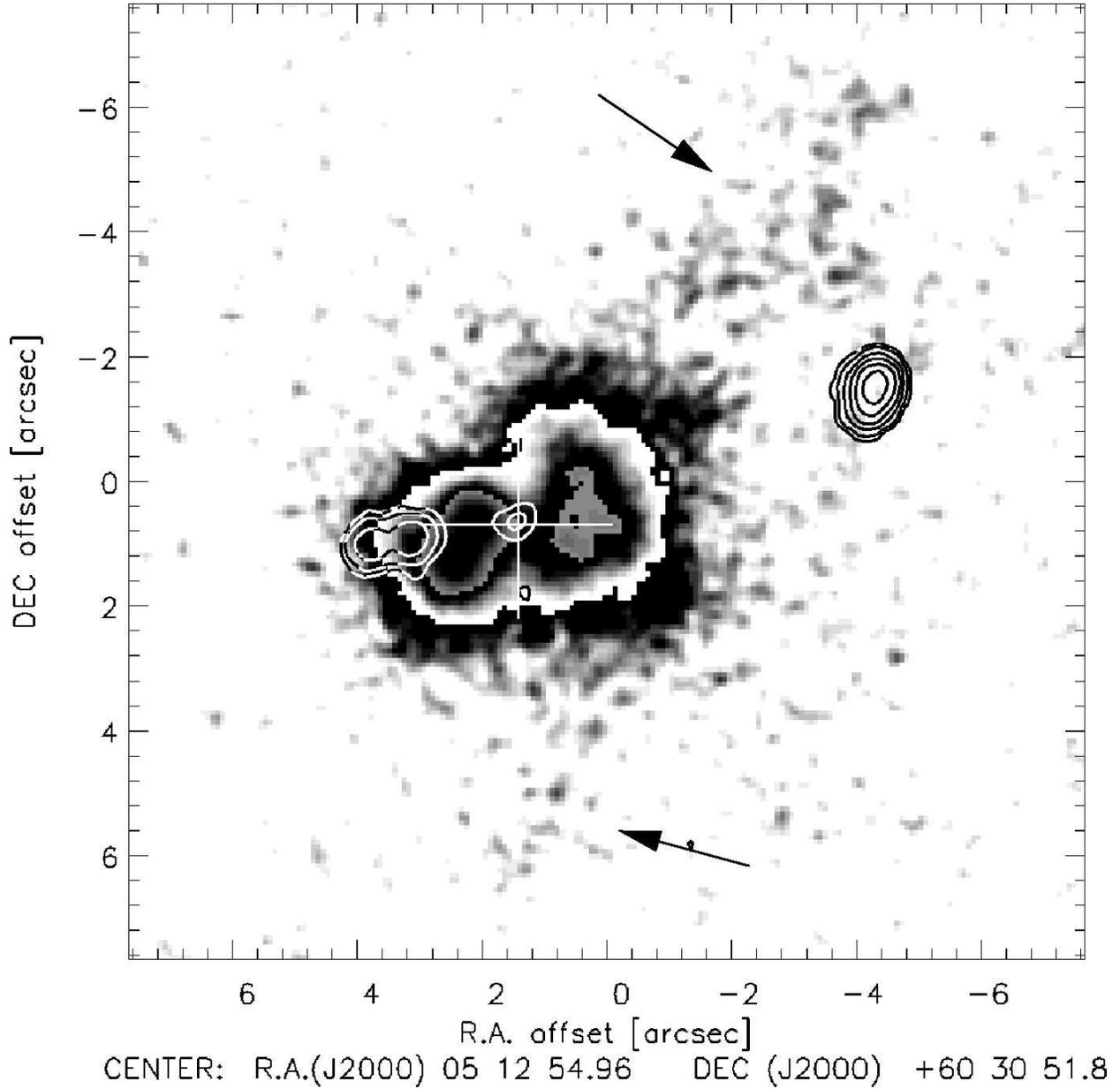}
\caption{ \lya\ image of \squ\ (grayscale) with 4.7\,GHz VLA map
\citep[contours;][]{Carilli97apj} overlaid. The radio core is
indicated with a cross, and the contour levels are 0.2, 0.4, 1.0, 2.0,
4.0 and 8.0 mJy beam$^{-1}$.  A very extended NW filament and
tentative evidence for a SE filament are indicated with
arrows.}\label{4c60Lya}
\end{figure}

\begin{figure}[t]
\plotone{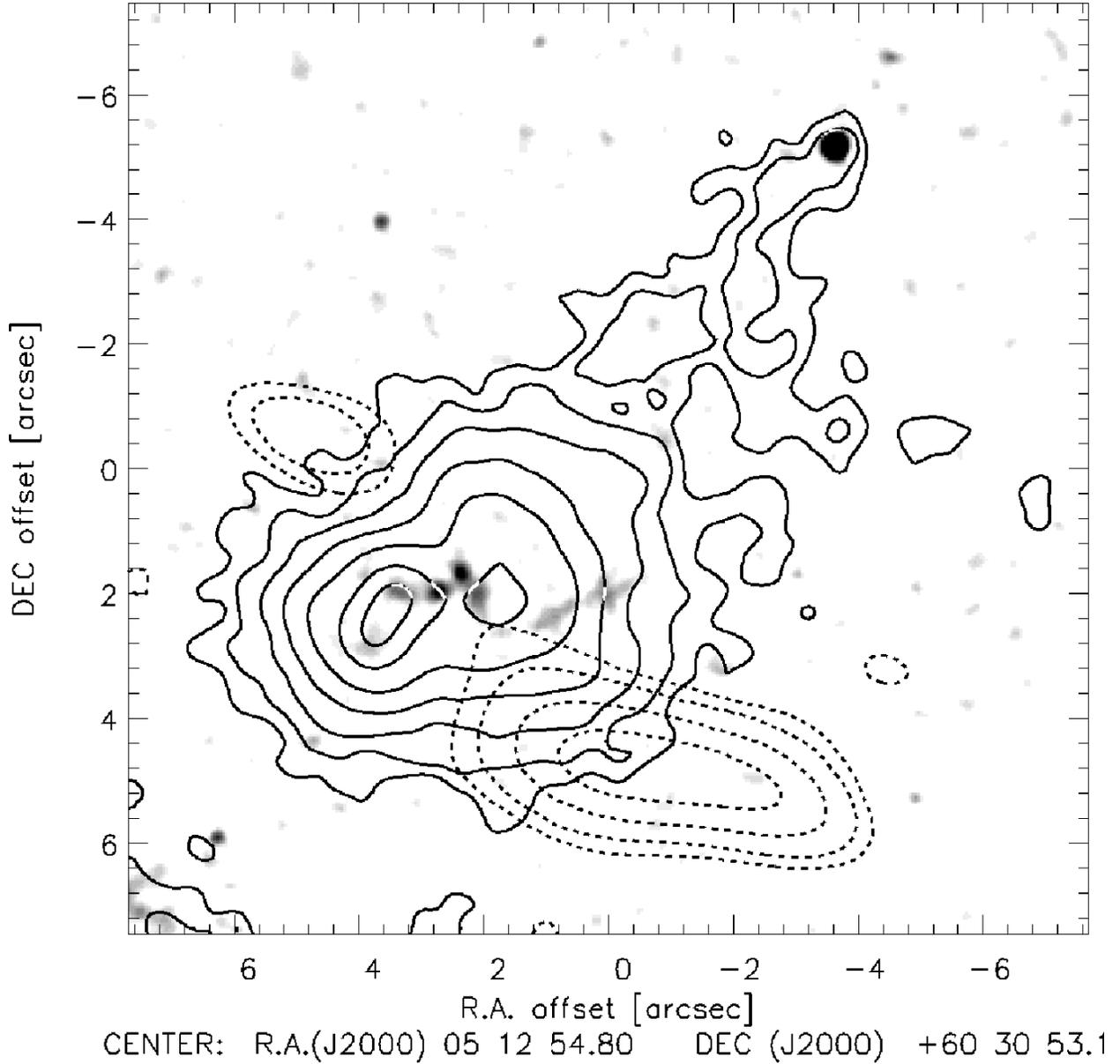}
\caption{HST F702W image of \squ\ with narrow-band image (solid
contours, not extinction corrected, including continuum) and 1.25\,mm
map \citep[dashed contours;][]{Papadopoulos00apj} overlaid.  The
narrow-band image has been heavily smoothed to bring out the low
surface brightness filament and contour levels are $6.7 \times
10^{-19} \times (5, 10, 20, 40, 80, 160, 240) \ergpspcm\,\rm
arcsec^{-2}$.  The levels for the 1.25\,mm emission are at 1.2, 1.6,
2.0, and 2.5\,mJy, with $\sigma = 0.6$\,mJy\,beam$^{-1}$, showing that
the NE component is a tentative detection comparable in size and shape
to the restoring beam. The galaxy at the tip of the narrow-band
filament is foreground at $z = 0.891$.}\label{4c60HST}
\end{figure}

\begin{figure}[t]
\plotone{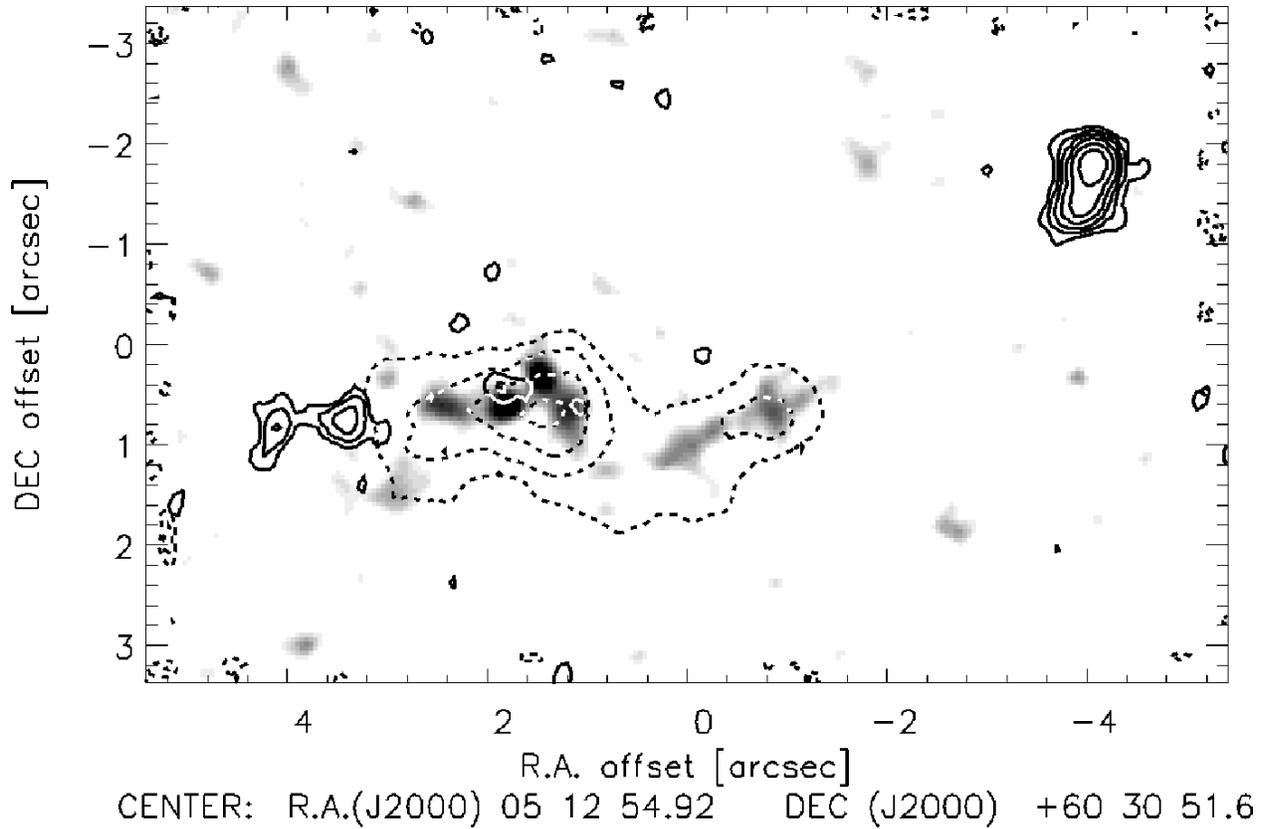}
\caption{Zoomed in grayscale representation of the HST F702W image of
\squ\ with 8.2\,GHz VLA map overlaid \citep[solid
contours;][]{Carilli97apj}.  The image has been smoothed to a
resolution of 0.25\arcsec, and ESI $R$-band contours (dashed) are
overlaid to better show the shape of the galaxy. The knotty, Z-shaped
structure of the aligned restframe UV continuum is apparent, while the
radio core might be coincident with the central gap between the two
brightest emission peaks. The radio contour levels are 0.08, 0.16,
0.3, 0.6, 1.2, and 2.4\,mJy\,beam$^{-1}$.}\label{4c60HSTzoom}
\end{figure}

\begin{figure}[t]
\plotone{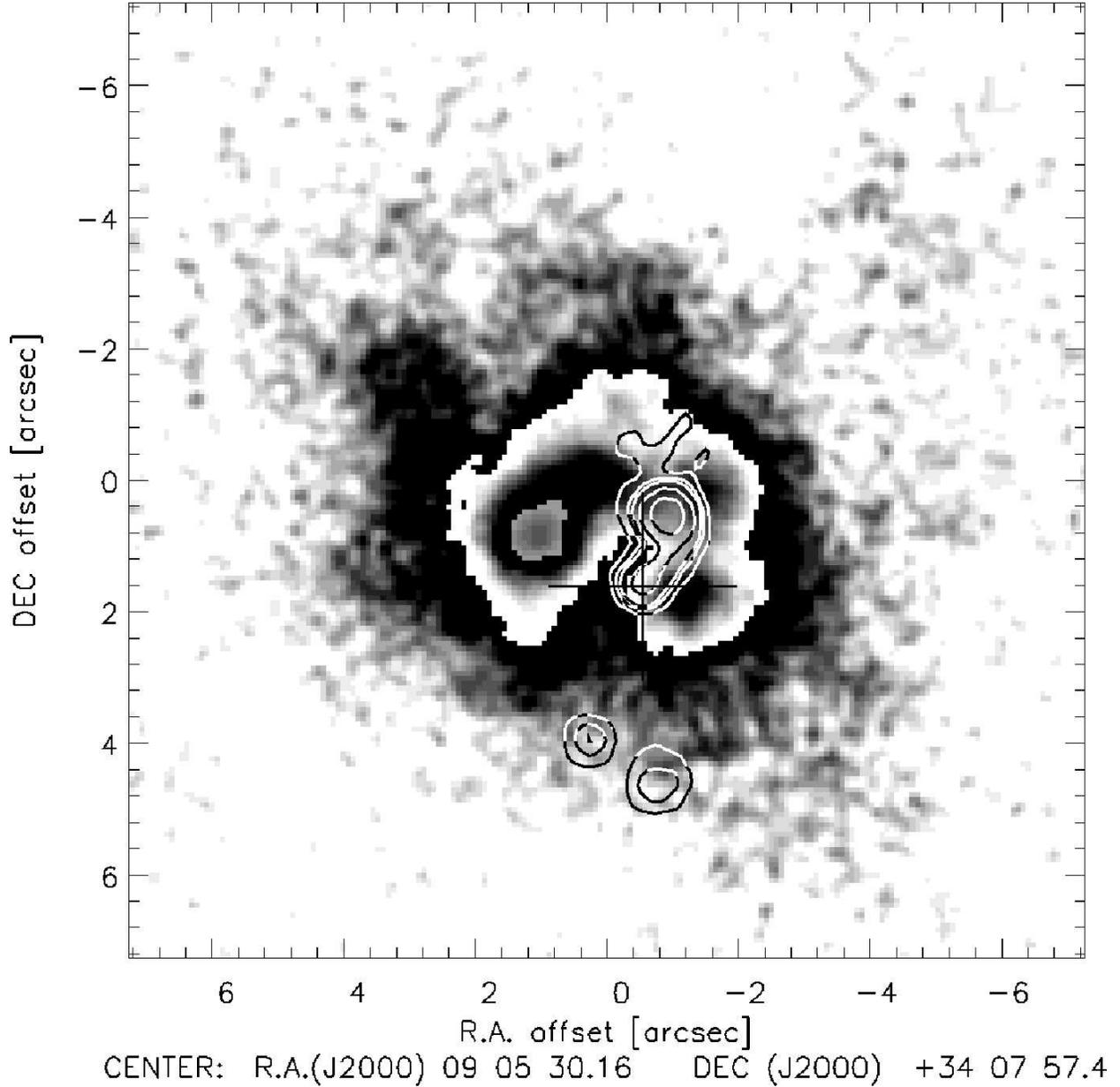}
\caption{A continuum-subtracted \lya\ image (grayscale) of \hor\ with
high resolution 4.9\,GHz VLA radio map
\citep[contours;][]{Carilli95aa} overlaid. The radio core is
identified with a cross, and the contour levels are 0.2, 0.8, 1.6,
6.4, and 25.6 mJy beam$^{-1}$.}\label{b20902Lya}
\end{figure}

\begin{figure}[t]
\plotone{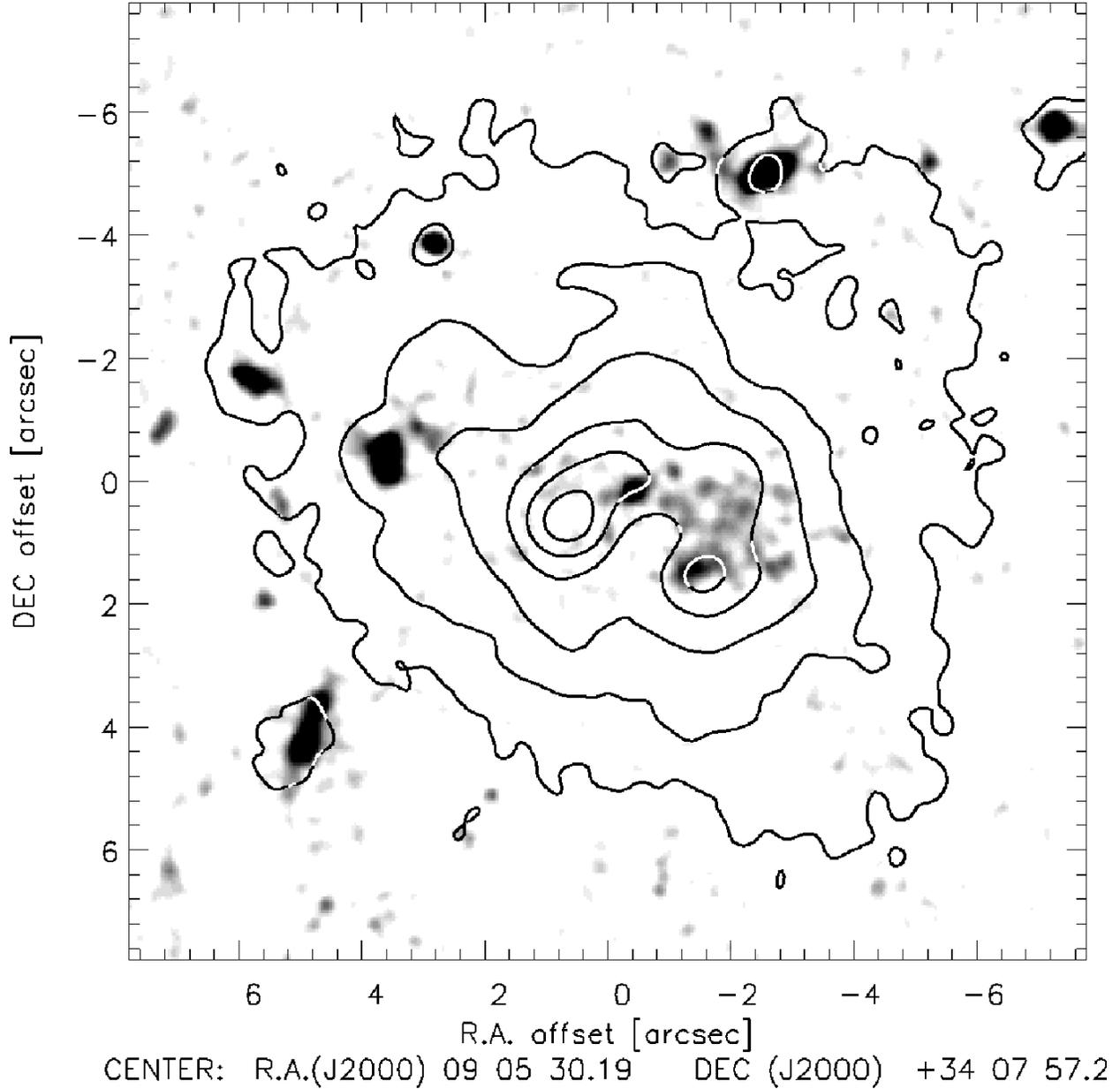}
\caption{Narrow-band image contours of \hor\ superposed on the HST
F622W image. The contour levels are $8.4 \times 10^{-19} \times (8,
20, 40, 80, 120, 160) \ergpspcm\,\rm arcsec^{-2}$, include continuum
emission, and have not been corrected for
extinction.}\label{b20902HST}
\end{figure}

In Abell~1795 there is a long, $\sim 80$\,kpc (projected size),
radially directed emission-line and X-ray filament associated with the
central cD galaxy.  \citet{Fabian01mnras} consider several possible
scenarios to explain this, including a ``contrail'' induced by ram
pressure from the radio source. However, based on kinematic and
cooling time considerations, and the fact that the cD galaxy is not
quite centered at the X-ray halo, Fabian \etal\ conclude that the
simplest explanation would be that the filament is a cooling wake
behind the galaxy as it moves within the X-ray halo. It is possible
that the bright, single filament in \squ\ might be explained in this
way but we consider this unlikely because of evidence for a very low
surface brightness cone-like structure (Fig. \ref{4c60Lya}).  The
multiple radially directed filaments in \tur\ would also be
inconsistent with such a model.

In NGC~1275 numerous radial emission-line filaments are found up to
$\sim 50$ kpc from the central AGN, with some tangential features at
large radii. These features overlap with a large radio halo centered on
the galaxy. The fairly constant surface brightness along the lengths
of the filaments and inferred physical parameters suggest that the
filaments formed as a result of compression of the intracluster gas by the
expanding radio source \citep{Conselice01aj}.  The correlation between
emission-line and radio source asymmetries in each of the three radio
galaxies shows that radio jets and lobes must indeed have a significant
impact on the ambient gas emissivity and emission-line morphology. 
Also, the eastern radio lobe of \tur\ appears to have associated
``plumes'' of enhanced emission straddling the lobe at both the SW and
SE, and the sizes of the radio structures are comparable to those of
the emission line gas. 

We note that even if the filaments are at a significant distance from
the observed radio source components, such as the SW lobe of \tur,
there still is a good reason to believe that they may be causally
related, in particular if there is supporting kinematic evidence
\citep[see 4C~29.30 for a nearby example;][]{vanBreugel86apj}. In the
canonical picture of radio sources the hotspots and lobes are
surrounded by bowshocks and cocoons of radio quiet, shock heated gas
with scale sizes wich are significantly larger than the observed radio
emission \citep{Carilli88apj,BegelmanCioffi89apj}. The emission-line
filaments are then located at the interface of the cocoons and the
ambient gas and are not in direct contact with the radio lobes or
hotspots and can even extend {\it beyond} radio hotspots
\citep[Figs. \ref{4c41Lya},\ref{4c60Lya};][]{Maxfield02aj} if one
accounts for projection effects and the fact that the radio
observations only show the highest surface brightness regions.

\subsection{Starburst superwinds}

Although \hzrgs\ may not resemble ``normal'' starburst systems in the
strictest sense, it seems reasonable to explore whether galactic
superwinds can explain the emission-line morphologies.
Observations of low redshift starburst galaxies show weakly collimated
bipolar outflows of gas with outflow velocities of several hundred
\kmps\ and up to scales of $\sim 10\kpc$
\citep{Heckman90apj}. H$\alpha$ emission line observations that trace
this gas show filaments and arclike structures which are possibly not
too dissimilar from the filaments that we can discern in the higher
redshift \lya\ nebulae.  Spectroscopic evidence for galactic outflows
has been found also at high redshifts
\citep[\eg][]{Pettini98apj,Pettini01apj,Dawson02apj} from metal
absorption lines that are blue-shifted by a few hundred to a few
thousand \kmps\ relative to the estimated systemic velocity of the
galaxies.  Recent observations of \ion{C}{4} absorptions systems along
the lines of sight to QSOs indicate enrichment of the intra-cluster
medium (ICM) even at redshifts higher than $z \sim 5$
\citep{Rauch01apj} which may have been caused by these outflows.

For a galactic outflow powered by a superwind to occur, the star
formation rate per unit area $\Sigma_{\rm SFR}$ must satisfy the
empirical relation $\Sigma_{\rm SFR} \gtrsim 0.1 \Msunpyr\kpc^{-2}$
\citep[\eg][]{Heckman01astroph}.  As we have argued above, based on
sub-mm detections and direct observations of stellar absorption lines,
\hzrgs\ are massive star forming systems, with SFRs of approximately
1000$-$2000\Msunpyr.  This implies SFR surface densities of
$\Sigma_{\rm SFR} \gtrsim 0.4 \Msunpyr\kpc^{-2}$ for the regions
corresponding to the extent of the $K$-band (restframe $B$) continuum.
Therefore, the minimum condition for an outflow can be satisfied
easily. Similarly, the SFR surface density based on the UV flux and
extent of the ``undisturbed'' southern region in \tur\ $\Sigma_{\rm
SFR} \gtrsim 33 \Msunpyr / 100 \kpc^2 \gtrsim 0.3 \Msunpyr\kpc^{-2}$
would also support a galactic wind.

Models show that starburst outflows fill an over-pressured cavity of
hot gas and expand into superbubbles, but they are usually not
energetic enough for the bubbles to burst and the gas to escape from
the host galaxy \citep{Heckman01astroph}.  This means that for the
massive \hzrgs\ this material is likely to reside at large radii but
still within the potential well of the galaxy.
\citet{TaniguchiShioya00apj} modelled the outflow of gas and find the
following relation for the radii of the shells:
$$r_{\rm shell} \sim 110 L_{\rm mech,43}^{0.2} n_{\rm H, -5}^{-0.2}
t_{8}^{0.6} \kpc,$$ with $L_{\rm mech} \sim 10^{43} \ergps \Msun^{-1}$
the mechanical luminosity released by supernovae in the starburst,
$n_{\rm H, -5}$ the hydrogen density in units of $10^{-5} \,\rm
cm^{-3}$ and $t_{8}$ the age of the starburst in units of $10^8$\,yr.
\citet{Dey97apj} modelled the stellar population of \tur\ with a
starburst of duration $1.6 \times 10^{7}$\,yr and an older population
of stars younger than $6 \times 10^{8}$\,yr. This gives a lower and
upperlimit for $t_{8}$ and \citet{Chambers90apj} estimated $n_{\rm H}
\sim n_{e} f_{\rm v}$ with $n_{e}$ the electron density and $f_{\rm
v}$ the volume filling factor of the clouds to be of order $20 \times 10^{-5}
\rm \,cm^{-3}$.  Applying these estimates to \tur\ we obtain: $20 \kpc
\leq r_{shell} \leq 170 \kpc$, resulting in a characteristic extent of
the superwind shells $l = 2r_{shell} \sim 150 \kpc$, consistent with the
sizes for the nebulae presented here.

\subsection{Radiation pressure driven outflows}

Recently, \citet{Dopita02astroph} presented a model to explain the
origin and kinematics of the narrow line regions in Seyfert
Galaxies. In this model ionized gas and dust are electrically locked
together and streaming from ionization fronts around photo-evaporating
clouds located in the ionization cones of the AGN. A similar process
might be envisaged in \hzrgs.  If the dust is destroyed somewhere
along the outflow, then the gas will continue to flow outward from the
cloud, but radiation pressure will no longer dominate the
dynamics. \citet{Aguirre01apj} also found that dusty outflows might be
radiation pressure driven and can reach distances of up to a few
100\,kpc.  These models suggest that radiative ablation by the central
source may be a significant if not dominant contributor to outflows.

Such a scenario would explain why the filaments are all oriented
radially, and why the focal point of the tails of the SW emission line
cloud in \tur\ seems to be the radio core while there is no obvious
connection with the radio emission.

\subsection{Obscured AGN}

For all three galaxies in the sample, and for 4C~23.56 which was
extensively studied by \citet{KnoppChambers97apj}, the radio core
appears to be located in a hole or at least a depression in the bright
central part of the \lya\ line emission. The best example of this is
\tur\ where the UV continuum also shows a gap of approximately
0\farcs4 ($\sim 3 \ph \kpc$).  Because the dips are also apparent in
the UV continuum and because there appears to be very bright emission
on either side, the gap must be due to an obscuring medium of high
optical depth.  \boa\ is the only \hzrg\ with a well studied \lya\
halo for which the radio core appears associated with a peak in \lya\
\citep{vanOjik96aa}. However, the central shape of \boa\ bears close
resemblance to the central cone shapes of \squ, and it is quite
possible that the central region would show two distinct components
with the radio core being in between if it were observed at higher
resolution.

\citet{Hippelein93nat} were the first to notice the gap in the \lya\
image of \tur\ and interpreted it as being due to a foreground
absorber being part of the \lya\ forest. The strong sub-mm dust
detection of \tur\ led \citet{Dunlop94nat} to speculate that this gap
might rather be due to a dusty disk. While this remains a viable
option, in the case of \squ\ which shows a similar UV continuum and
emission line dip, sub-mm observations show that {\it most} of the
dust is located ``outside'' the galaxy.

At least two of the three radio galaxies that we present here, show a
distinct biconical shape in the brightest parts of their nebulosities.
This morphology suggests that the nucleus contributes significantly to
the photoionization, and is surrounded by an obscuring torus which we
observe under an angle as favored by the quasar-radio galaxy
unification scheme \citep{Barthel89apj}.  This putative torus would be
much larger than what is traditionally assumed and it would be more
appropriate to refer to this feature as a dust lane. Dust lanes have
been studied in nearby FR I radio galaxies \citep{Verdoes99aj}, and
appear to be orthogonal to the radio axis independent of the
orientation of the galactic disk. Verdoes \etal\ found that these
features typically range from 200\,pc to 4.5\kpc, while features
larger than $\sim 1\kpc$ show peculiar morphologies and are possibly
still settling towards stable orbits.  This suggests that the large
sizes seen in \hzrgs\ might be the result of the availability of
substantial amounts of debris and that the torus/dust lane is still
forming and has not yet reached its final configuration.

\section{Conclusions}

We have presented a morphological study of the \lya\ emission line
nebulae of three radio galaxies at redshifts $z \sim 3.4 - 3.8$.

Our new findings are that the emission line nebulae show giant, 100 -
200 kpc low surface brightness emission and large (up to $80\kpc$)
filamentary and cone-shaped structures, which are even more extended
than known previously. We have also found indirect evidence for extended
\lya\ absorption due to neighboring galaxies in the local foreground to
the \tur\ nebula and, in all three cases, that the AGN appear obscured
by surrounding dust. 

In order to explain these observations, we have presented a scenario
in which primordial cooling flows in Cold Dark Matter halos form
multiple stellar systems which merge, building the stellar bulges of
future massive galaxies and triggering (radio-loud) SMBH activity.
The radio jets, starburst superwinds and radiative ablation from the
AGN may all contribute to cause the observed radially directed
filamentary structures, in at least two of the three objects studied,
and may provide a feedback mechanism which regulates stellar bulge and
SMBH growth. To study these processes will require further
observations of the physical properties of the gas, and numerical
simulations of the effects of jets and AGN radiation on dusty,
astrophysical plasmas. Some regions of the outermost halos appear
unaffected by any radio/AGN activity. This suggests that this gas must
either be infalling for the first time or be a remant of previous
outflows.

Deep spectroscopy targeting non-resonant emission lines are important
to constrain the various scenarios, and determine the gas kinematics,
metallicities and sources of ionization.  The observations presented
here also show that filamentary structure is not always aligned with
the radio axis, and that long slit spectroscopy can easily miss large,
bright filaments.  Therefore, to fully exploit narrow-band
emission-line images as a tool for studies of galaxy formation
requires the use of ``3-Dimensional'' (2-D spatial + 1-D spectral)
imaging devices on large telescopes
\citep{vanBreugelBlandHawthorn00pasp}.  Several such devices are
currently being build, including the LLNL Imaging Fourier Transform
Spectrometer (Wurtz \etal 2001), and the OSIRIS imaging spectrograph
for the Spanish 10.4\,m telescope GTC at La Palma (Cepa \etal 2001).

\vskip 0.3cm

It is a pleasure to thank Chris Carilli for making available the VLA
radio maps, Padeli Papadopoulos for providing us with the PdB images,
Bj\"orn Heijligers for IDL tips and tricks, and the staff at W.M. Keck
Observatory for their efficient assistance.  We thank the referee,
Montse Villar-Mart{\'\i}n for many helpful suggestions that improved
the manuscript. The work of M.R., W.d.V., W.v.B., A.S., and M.L. was
performed under the auspices of the U.S. Department of Energy,
National Nuclear Security Administration by the University of
California, Lawrence Livermore National Laboratory under contract
No. W-7405-Eng-48. W.v.B. also acknowledges NASA grants GO 5429, 5765,
5940, 6608, and 8183 in support of \hzrg\ research with HST at
LLNL. M.L. performed part of this work at the Jet Propulsion
Laboratory, California Institute of Technology, under contract with
NASA.  M.D. acknowledges the support of the ANU and the Australian
Research Council (ARC) for his ARC Australian Federation Fellowship,
and also under the ARC Discovery project DP0208445.  This research has
made use of the USNOFS Image and Catalogue Archive operated by the
United States Naval Observatory, Flagstaff Station
(http://www.nofs.navy.mil/data/fchpix/), and the NASA/IPAC
Extragalactic Database (NED) which is operated by the Jet Propulsion
Laboratory, California Institute of Technology, under contract with
the National Aeronautics and Space Administration.

\clearpage

\begin{deluxetable}{lcccccccccc}
\tabletypesize{\scriptsize}
\tablewidth{460pt}
\tablecaption{Summary of Keck observations\label{obstable}}
\tablehead{
\colhead{}
& \multicolumn{2}{c}{Coordinates \tablenotemark{a}}
& \colhead{}
& \multicolumn{4}{c}{Narrow-band}
& \colhead{}
& \multicolumn{2}{c}{R-band} \\
\cline{2-3} \cline{5-8} \cline{10-11} \\
  \colhead{Object}
& \colhead{RA}
& \colhead{DEC}
& \colhead{}
& \colhead{Gal.Ext. \tablenotemark{b}}
& \colhead{}
& \colhead{t$_{\rm obs}$}
& \colhead{1$\sigma$ det.surf.br. \tablenotemark{c}}
& \colhead{}
& \colhead{t$_{\rm obs}$}
& \colhead{1$\sigma$ det.surf.br. \tablenotemark{c}}
\\
& \colhead{(J2000)}
& \colhead{(J2000)}
& \colhead{}
& \colhead{mag}
& \colhead{}
& \colhead{(ks)}
& \colhead{(AB mag)}
& \colhead{}
& \colhead{(ks)}
& \colhead{(AB mag)}
} 
\startdata
\hor & 09 05 30.11 & 34 07 55.9 & & 0.101 & & \phantom{0}9.6 &  29.04, 27.4 &  & 2.4 & 29.70, 28.3 \\
\squ & 05 12 55.17 & 60 30 51.1 & & 1.609 & & \phantom{0}7.2 &  28.85, 27.3 &  & 2.0 & 29.64, 28.1  \\
\tur & 06 50 52.14 & 41 30 30.7 & & 0.393 & & 27.6 & 29.85, 28.1 &  & 4.0  & 30.08, 28.7 \\
\enddata
\tablenotetext{a}{Position of the radio core
\citep{Carilli94aj,Carilli97apj,Carilli95aa}}
\tablenotetext{b}{Galactic extinction at the central wavelength of the
narrow-band filters}
\tablenotetext{c}{Formal 1$\sigma$ detection limit: in the seeing
disc and in a 3\arcsec\ diameter aperture respectively}
\end{deluxetable}

\clearpage

\begin{deluxetable}{lccccccccc}
\tabletypesize{\scriptsize}
\tablewidth{500pt}
\tablecaption{Properties of the \lya\ nebulae\label{proptable}}
\tablehead{\colhead{Object} & \colhead{$z$} & \colhead{NB \tablenotemark{a}} & \colhead{BB$-$NB}
& \colhead{F$_{\rm Ly\alpha}$ \tablenotemark{b}}
& \colhead{log L$_{\rm Ly\alpha}$ \tablenotemark{c}}  & \colhead{Extent} & \colhead{Size \tablenotemark{c}}
& \colhead{Mass \tablenotemark{d}} & \colhead{$SFR$ \tablenotemark{e}}  \\
& & \colhead{mag} & \colhead{mag} & \colhead{\ergpspcm} & \colhead{\ergps} & \colhead{\arcsec\ $\times$ \arcsec}& \colhead{\kpc\ $\times$ \kpc} & \colhead{$10^{8}$\,\Msun} & \colhead{\Msunpyr} }
\startdata
\hor & 3.395 & 19.20 $\pm$ 0.004 & 2.81 & $4.9 \times 10^{-15}$ & 44.77 & 10 $\times$ 8\phantom{0} &
$\phantom{1}80 \times \phantom{1}64$ & \phantom{1}30$f_{\rm v}^{-\case{1}{2}}$ & $> \phantom{1}450 f_{\rm esc}^{-1}$ \\
\squ & 3.791 & 19.97 $\pm$ 0.050 & 3.26 & $8.8 \times 10^{-15}$ & 45.14 & \phantom{0}9 $\times$ 7\phantom{0}& $\phantom{1}68 \times \phantom{1}53$ & \phantom{1}30$f_{\rm v}^{-\case{1}{2}}$ & $> 1100 f_{\rm esc}^{-1}$ \\
\tur & 3.798 & 18.76 $\pm$ 0.001 & 3.38 & $9.0 \times 10^{-15}$ & 45.15 & 25 $\times$ 17 & $190 \times 130$ & 130$f_{\rm v}^{-\case{1}{2}}$  & $> 1100 f_{\rm esc}^{-1}$\\
\enddata \tablenotetext{a}{Determined using SExtractor
\citep{BertinArnouts96aa} with automatic aperture selection and
deblend parameter optimized for detecting large structures. The
uncertainties quoted are the formal error estimates given by
SExtractor. The magnitudes were also determined in IRAF using a curve
of growth for the cumulative magnitude in increasing radii which
resulted in magnitudes differing by at most 0.3 mag.}
\tablenotetext{b}{Corrected for Galactic extinction}
\tablenotetext{c}{Assuming a cosmology with $\OmM = 0.3,
\OmL = 0.7$, and $H_{0} = 65\kmps\, \rm Mpc^{-1}$}
\tablenotetext{d}{These mass estimates are derived from $M =
n_{e} m_{p} f_{\rm v} V$ and $L_{\rm Ly\alpha} = 4 \times 10^{-24}
n_{e}^{2} f_{\rm v} V \ergps$ \citep[\eg][]{McCarthy90apj} assuming
cylindrical symmetry and case B recombination. We have assumed a filling
factor $f_{\rm v} = 10^{-5}$, which is extremely uncertain
\citep{Heckman82apj}.}
\tablenotetext{e}{The estimated $SFR \geq 8.12 \times 10^{-43}
f_{\rm esc}^{-1} L_{\lya}$ is very uncertain and based on the
assumption that all of the \lya\ flux is the result of Case B
recombination of photoionization by a young stellar population (see
Section 3.1).}
\end{deluxetable}

\clearpage

\clearpage

\clearpage

\clearpage

\clearpage

\clearpage

\clearpage

\clearpage

\clearpage

\end{document}